\documentclass[letterpaper, onecolumn, 10pt]{extarticle}

\usepackage{titling}
\usepackage[english]{babel}
\usepackage[utf8]{inputenc}
\usepackage[T1]{fontenc}
\usepackage{lmodern,textcomp}
\usepackage[left=1in, right=1in, top=0.5in, footskip=0.5in]{geometry}
\usepackage{graphicx}
\usepackage[space]{grffile}
\usepackage{epstopdf}
\usepackage{wrapfig}
\usepackage{multirow}
\usepackage{listings}
\usepackage{setspace}
\usepackage{indentfirst}
\usepackage{rotating}
\usepackage{hyperref}
\hypersetup{
    colorlinks=true, 
    linktoc=all,     
    linkcolor=blue,  
    allcolors=blue,
}
\usepackage{nameref}
\usepackage{mathtools}
\usepackage{amsmath,amsthm}
\usepackage[makeroom]{cancel}
\usepackage{amsfonts}
\usepackage{amssymb}
\usepackage[font=small,labelfont=bf,format=plain,justification=justified]{caption}
\usepackage{floatrow}
\usepackage{subcaption}
\usepackage{array}
\newcolumntype{H}{>{\setbox0=\hbox\bgroup}c<{\egroup}@{}}
\usepackage{pstricks}
\usepackage{epic}
\usepackage{enumerate}
\usepackage{booktabs}
\usepackage{longtable}
\usepackage[square,numbers,comma,sort&compress]{natbib}
\usepackage{authblk}
\usepackage{ulem} 
\usepackage{siunitx}
\usepackage{tabularx} 
\usepackage{booktabs}

\definecolor{forestgreen}{rgb}{0.33,0.61,0.34}

\graphicspath{{./figures/}}

\title{Identifying key players in dark web marketplaces}

\author[1,2]{Elohim Fonseca dos Reis}
\author[3]{Alexander Teytelboym}
\author[4]{Abeer ElBahrawy}
\author[5]{Ignacio De Loizaga}
\author[1,2,6,*]{Andrea Baronchelli}
\affil[1]{The Alan Turing Institute, London NW1 2DB, UK}
\affil[2]{Department of Mathematics, City, University of London, London EC1V 0HB, UK}
\affil[3]{Department of Economics, University of Oxford, Oxford OX1 3UQ, UK}
\affil[4]{Chainalysis Inc, New York, NY, USA}
\affil[5]{PayPal Inc, San Jose, CA, USA}
\affil[6]{UCL Centre for Blockchain Technologies, University College London, London WC1E 6BT, UK}
\affil[*]{Corresponding author: abaronchelli@turing.ac.uk}
\date{}                     
\begin{document}
\maketitle

\begin{abstract}
Dark web marketplaces have been a significant outlet for illicit trade, serving millions of users worldwide for over a decade. However, not all users are the same. 
This paper aims to identify the key players in Bitcoin transaction networks linked to dark markets and assess their role by analysing a dataset of 40 million Bitcoin transactions involving 31 markets in the period 2011-2021. 
First, we propose an algorithm that categorizes users either as buyers or sellers and shows that a large fraction of the traded volume is concentrated in a small group of elite market participants. 
Then, we investigate both market star-graphs and user-to-user networks and highlight the importance of a new class of users, namely `multihomers' who operate on multiple marketplaces concurrently. 
Specifically, we show how the networks of multihomers and seller-to-seller interactions can shed light on the resilience of the dark market ecosystem against external shocks. 
Our findings suggest that understanding the behavior of key players in dark web marketplaces is critical to effectively disrupting illegal activities.
\end{abstract}

\section*{Introduction} \label{sec:introduction}

The dark web has been home of many unregulated online commercial platforms facilitating the trade of illicit goods \cite{EuropolEMCDDA2017DrugsAndDarknet, Gwern.net, EMCDDA2019DrugSupplyDarknet, EMCDDA2019EUDrugMarkets, EMCDDA2020CovidAndDrugs, Interpol2020CombCyberCrime, Europol2021IOCTA, Chainalysis2021CrimeReport, Chainalysis2022CrimeReport, Chainalysis2023CrimeReport}.
This ecosystem, composed of the dark web marketplaces (DWMs) and the network of user-to-user (U2U) transactions \cite{Barratt2016IntJDrugPolicy, Munksgaard2020TrendsCrime, Nadini2022SciRep}, has proven to be extremely reactive to the demands of society and resilient against external shocks \cite{EMCDDA2020CovidAndDrugs, Bracci2022PNASNexus, Europol2021IOCTA, Abeer2020SciRep}.
Despite the risks associated with their illegal nature, millions of users have traded on these platforms since the launch of Silk Road, the first modern DWM \cite{Christin2013IntConfWWW}.
The popularity of DWMs stems from users being able to access them easily and anonymously, and trade items that are not available in regulated markets.
Owing to their unregulated character, DWMs offer no protection to buyers and sellers.
Many DWMs were closed, either by law enforcement operations or by exit scams, leaving their users with significant losses  \cite{EuropolEMCDDA2017DrugsAndDarknet}.
This uncertainty has not prevented this ecosystem from growing, resilient to external shocks thanks to a `Whack-a-Mole' dynamics: when a market is closed, users migrate to an alternative platform in a swift and coordinated fashion \cite{Abeer2020SciRep, Europol2021IOCTA}.

Surprisingly, although DMWs have gained significant attention from the scientific community and law enforcement agencies, little is known about the key players sustaining this unique adaptability and elastic dynamics. 
In particular, DWM transaction networks have been investigated so far mainly with focus on DWM \textit{users}, without distinguishing between buyers and sellers, and neglecting the different weight that more active users may have in the system.
The reason is that the operational structure of DWMs inherently hides the seller-buyer link, as all transactions are made through the marketplace. 
Buyers send money to the marketplace, which in turn sends the money to the seller.
Thus, for example, while previous studies based on scraped listings and surveys have revealed that buyers and sellers respond differently to a closure \cite{Zambiasi2022JEconBehavOrgand, Galenianos2012RevEconStud}, further analyses in this direction have been hindered by the lack of heuristics able to identify these two key classes of actors in transaction networks.

Here, we set out to find the main actors in the DWM ecosystem and assess their systemic impact on a dataset of 40 million Bitcoin transactions involving 31 markets in the period 2011-2021. First, we propose a simple algorithm to identify buyers and sellers. Importantly, the algorithm returns reasonable estimates for the number of sellers when compared against a benchmark of nine DWMs where official estimates exist. Then, we reveal a concentration of activity around an elite group of participants, where a large fraction of the traded volume is driven by a small number of players. By formally describing the ecosystem of DWMs as a temporal network where nodes are the entities and directed edges are transactions pointing from source to destination, we consider different networks of buyers and sellers, promoting different functions in the ecosystem.
In particular, we analyse the networks of `multihomers', defined as users that are simultaneously trading in multiple markets.
We show that these users play a crucial role in the connectivity of the ecosystem because they act as connectors between markets.
Analogously, we identify and characterise `multisellers' (i.e., multihomers that are sellers) and `multibuyers' (i.e., multihomers that are buyers).
Furthermore, we analyse the seller-to-seller (S2S) network, i.e., the network composed only of transactions among sellers, which can be regarded as a supply chain network of illicit goods and services.
We highlight that these networks exhibit different resilience regimes in the presence of external shocks, the ecosystem's resilience being mostly guaranteed by the network of buyers rather than sellers.

\section*{Results} \label{sec:results}

\subsection*{Buyers and sellers}

\begin{figure*}[ht]
	\centering
	\includegraphics[scale=0.45]{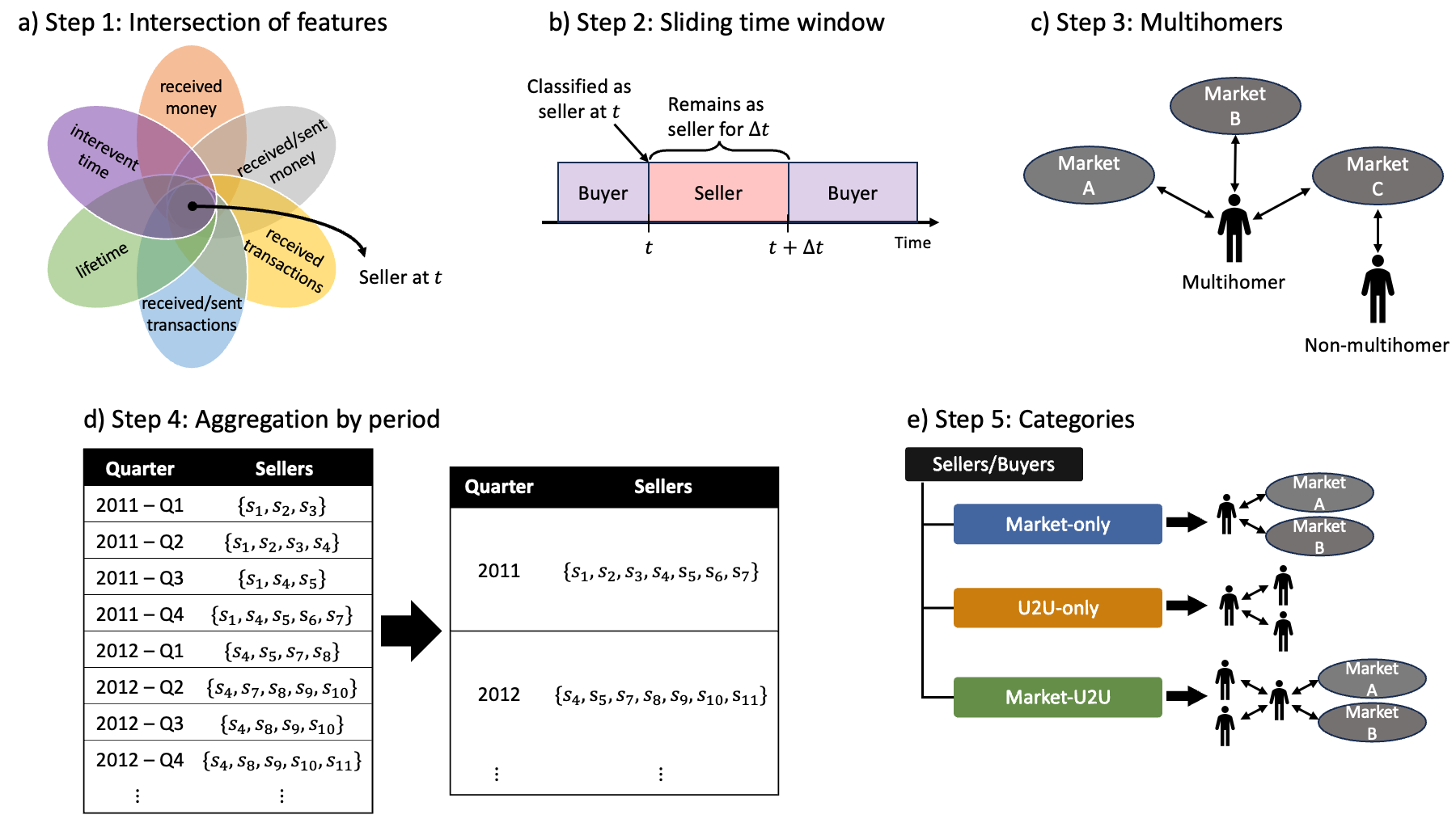}
	\caption{
        \textbf{The five steps of the classification of entities as buyer and sellers.}
        a) Step 1: Intersection of features -- an entity is classified as a seller at time $t$ when they satisfy the six features simultaneously.
        b) Step 2: Sliding time window -- when an entity is classified as a seller, they remain as a seller during $\Delta t$ days.
        c) Step 3: Multihomers -- we identify sellers and buyers that are simultaneously active in more than one market.
        d) Step 4: Aggregation by period -- we aggregate the lists of obtained sellers and buyers according to a given calendar period.
        e) Step 5: Categories -- buyers and sellers are divided into the three categories.
		}
	\label{fig:classification_steps}
\end{figure*}

To characterize the structure and dynamics of the ecosystem of DWMs, we start by classifying all traders either as sellers or buyers.
We implement a method of classification based on money exchanged, number of transactions, and time activity for each entity, as illustrated in Fig.~\ref{fig:classification_steps} (see Methods for details).
We consider each market separately, i.e., we obtain a time series of sellers for each market, and we use the same method and classification parameters in the U2U network.
Therefore, an entity can be classified as a seller in one or more markets and/or the U2U network simultaneously.
We feed the classification process with seven parameters: 
minimum money received (denoted by $M$), 
minimum ratio between received and sent money (denoted by $\alpha$), 
minimum number of transactions (denoted by $T$), 
minimum ratio between number of received and sent transactions (denoted by $\beta$), 
minimum lifetime (denoted by $L$), 
maximum mean interevent time of transactions (denoted by $\tau$),
and a sliding time window (denoted by $\Delta t$).

We set $M = 100$ USD, $\alpha = 3$, $T = 10$, $\beta = 3$, $L = 10$ days, $\tau = 10$ days, and $\Delta t = 30$ days in the following.
The classification method is considerably robust with respect to the values assigned to the classification parameters (see Supplementary Information).

\begin{figure}[ht]
    \includegraphics[scale=1]{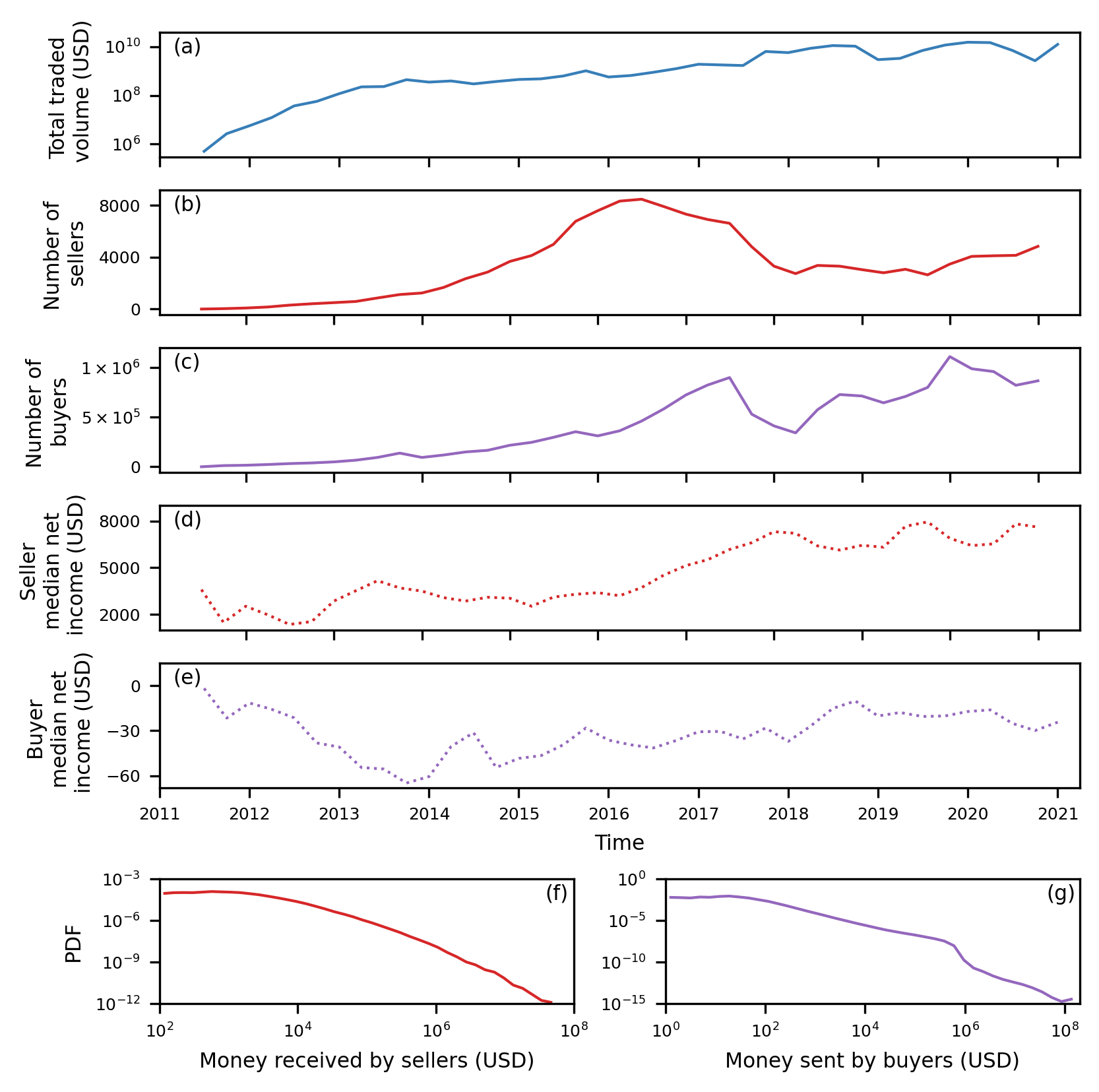}
	\caption{
		\textbf{Evolution of buyers and sellers in the ecosystem of DWMs.} 
		We consider the whole ecosystem, i.e., all markets and the U2U network.
        In panel (a), we show the total quarterly traded volume in USD. 
		In panels (b) and (c), we show the number of all sellers and buyers per quarter, respectively.
		In panels (c) and (d), we show the median net income in USD of all sellers and buyers per quarter, respectively.
        In panels (e) and (f), we show the PDF of the total money received by each seller and the total money sent by each buyer, respectively.}
	\label{fig:overall_numbers_PDF_traded_vol_s30}
\end{figure}

The evolution of the ecosystem of all buyers and sellers obtained from the considered markets and the U2U network is shown in Fig.~\ref{fig:overall_numbers_PDF_traded_vol_s30}.
The total quarterly traded volume is shown in Fig.~\ref{fig:overall_numbers_PDF_traded_vol_s30}(a).
Although it shows fluctuations, including those caused by external shocks, the ecosystem is overall growing, in terms of traded volume.

Our classification shows that the number of sellers is significantly smaller than the number of buyers, as shown in Figs.~\ref{fig:overall_numbers_PDF_traded_vol_s30}(b) and \ref{fig:overall_numbers_PDF_traded_vol_s30}(c), respectively.
The number of actors in the ecosystem is affected by several factors, especially market closures.
Notably, the number of buyers and sellers significantly drops after the operation Bayonet in the last quarter of 2017, which shut down AlphaBay and Hansa markets, causing a major shock in the ecosystem \cite{EuropolBayonet}.
However, the number of buyers rapidly recovers its previous values, which does not happen to sellers.

To compare the financial behaviour between sellers and buyers, we compute the median net income, defined as the median of the difference between the cumulative money received and sent from all transactions (in all markets and U2U network) made by each entity as a function of time, as shown in Figs.~\ref{fig:overall_numbers_PDF_traded_vol_s30}(d) and \ref{fig:overall_numbers_PDF_traded_vol_s30}(e).
The median net income is positive for sellers while negative for buyers throughout the whole period of observation.
This result is not trivial because, although the classification induces a positive net income for sellers, it is performed on each market and the U2U network separately, while the median net income is computed based on all transactions made by an entity.
In fact, when we compute the total net income for each seller, a considerable fraction (16\%) has a negative net income because they spend in markets where they are not classified as sellers, or in the U2U network.
When we aggregate all entities that have been classified as seller, we find that 5\% of sellers -- those who have received more than $2.3 \times 10^5$ USD -- account for 81\% of the total amount of money received by all sellers.
Conversely, we find that 5\% of buyers -- those who have sent more than $2.3 \times 10^4$ USD --  account for 92\% of the total amount of money sent by all buyers.
Therefore, there is small fraction of actors responsible for moving most of the traded volume in both directions, i.e., buying and selling.
We observe this concentration of traded volume in the probability density functions (PDFs) of the total money received by each seller and the total money sent by each buyer, as shown in Figs.~\ref{fig:overall_numbers_PDF_traded_vol_s30}(f) and \ref{fig:overall_numbers_PDF_traded_vol_s30}(g), respectively.
In both cases, we see a significant heterogeneous distribution.


\subsection*{Multihomers and categories} \label{sec:multihomers_categories}

In order to analyse the connectivity of the whole ecosystem, i.e., how markets are connected with each other, we consider sellers and buyers that are simultaneously active on multiple platforms. These multihomers, i.e., multisellers and multibuyers, play the role of edges between markets.

Additionally, three mutually exclusive categories of sellers (buyers) naturally emerge:
\begin{enumerate}
\item Market-only sellers (buyers), who are active in one or more markets and not in the U2U network;  
\item U2U-only sellers (buyers), who are active only in the U2U network, and
\item Market-U2U sellers (buyers), who are simultaneously active in both markets and the U2U network.
\end{enumerate}

\begin{figure*} [ht]
	\centering
	\includegraphics[scale=1]{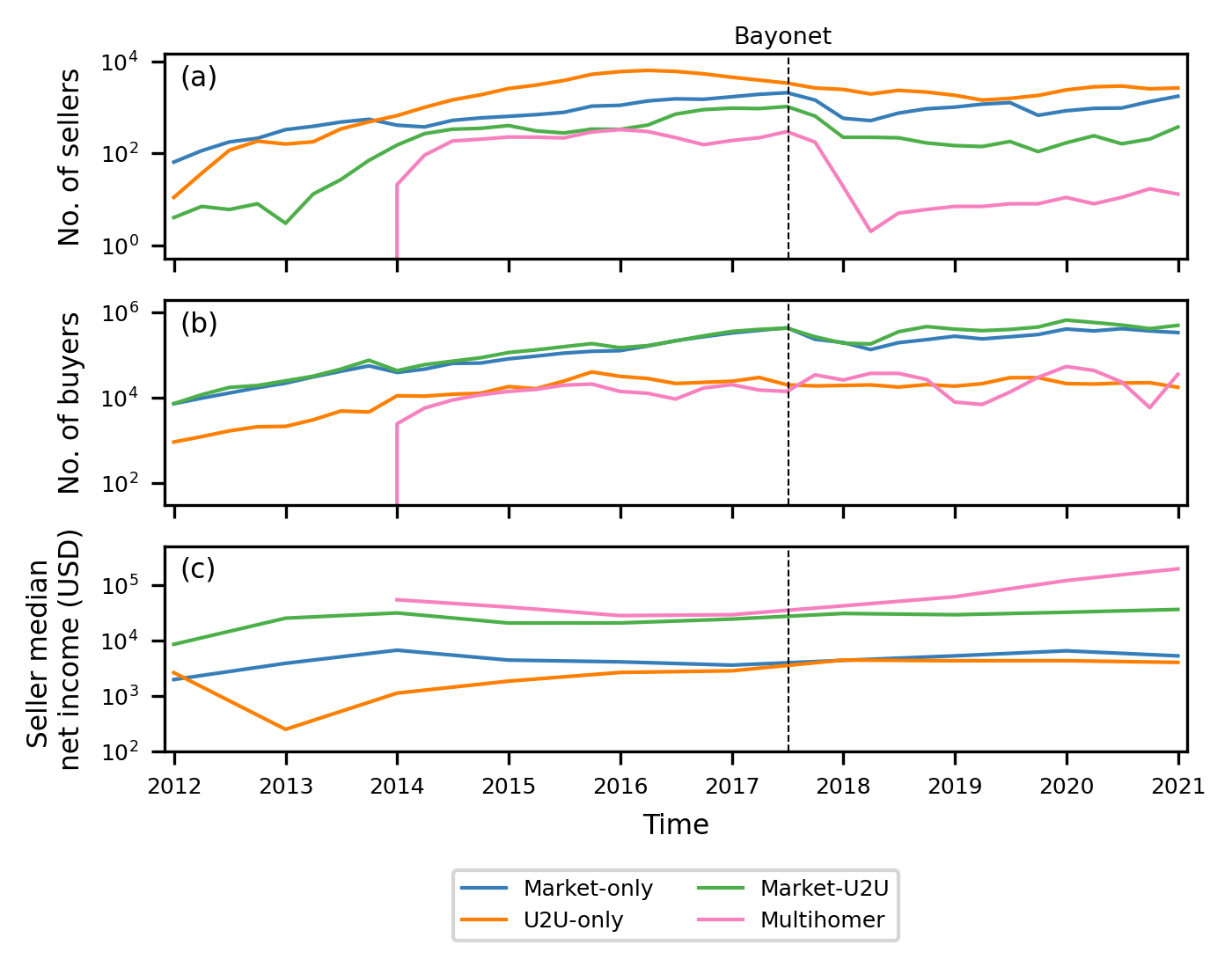}
	\caption{
		\textbf{The evolution of the different types of sellers and buyers.}
       		(a) The number of sellers for each category and multisellers as a function of time. 
	        (b) The number of buyers for each category and multibuyers as a function of time.
            (c) The yearly median net income of sellers.
		The dashed vertical line shows the time of the major external shock caused by operation Bayonet.
		}
	\label{fig:categories}
\end{figure*}

The number of sellers in each category and multisellers as a function of time is shown in Fig.~\ref{fig:categories}(a).
Until the end of 2013, when Silk Road is the dominant market (and the only one during most of this period), market-only sellers is the dominant category, and there are no multisellers.
From the last quarter of 2013, U2U-only sellers become the dominant category of sellers and remains dominant throughout the rest of the observation period.
The larger number of U2U-only sellers is in accordance with previous results that showed that the traded volume in the U2U network is larger than that of DWMs \cite{Nadini2022SciRep}.

\newpage

With the advent of several markets at the beginning of 2014, the number of multisellers rapidly grows, representing more than 20\% of all sellers until the beginning of 2016 (see Supplementary Information).
During 2016 and 2017, AphaBay becomes the dominant market, polarizing sellers around its own ecosystem, such that the fraction of multisellers decreases to 10\% of all sellers until its closure.
Then, after operation Bayonet, the number of sellers in all categories significantly drops, as shown in Fig.~\ref{fig:categories}(a).
Notably, the number of multisellers suffers the largest drop of $-97\%$.
Moreover, while the other categories of sellers recover their previous numbers, the number of multisellers remains small after that shock.

The scenario buyer scenario is different, as shown in Fig.~\ref{fig:categories}(b).
Throughout the whole period of observation, the dominant category of buyers is market-U2U buyers followed by market-only buyers, representing on average 52\% and 42\% of all buyers, respectively.
The U2U-only category is comparatively small, representing only 6\% of all buyers on average.
The number of market-U2U and market-only buyers also drops as a consequence of operation Bayonet.
However, compared to sellers, the drop is notably smaller, and the number of buyers rapidly recovers to previous values.
On the other hand, the number of U2U-only buyers is less affected.
Moreover, the number of multibuyers increases, which suggests a fast response from buyers to external shocks by trying to diversify their sources.

In Figure \ref{fig:categories}(c) we compare the yearly median net income of the categories of sellers.
Multisellers have the largest net income consistently throughout the period of observation.
Market-U2U sellers follow them, then market-only sellers, and finally U2U-only sellers.
Therefore, although larger in number, U2U-only sellers typically make the smallest net income.
This suggests that sellers with more diverse sources of income, such as multisellers and market-U2U sellers, are able to produce a higher income.
Moreover, the median net income of all types of sellers does not seem to be affected by any external shocks.

The observed changes in the number of sellers and buyers shown in Fig.\ref{fig:categories} are reflected in the networks where multihomers are regarded as edges connecting markets.

\subsection*{Multiseller network} \label{sec:multiseller_network}

\begin{figure*}[ht!]
	\centering
	\includegraphics[scale=1]{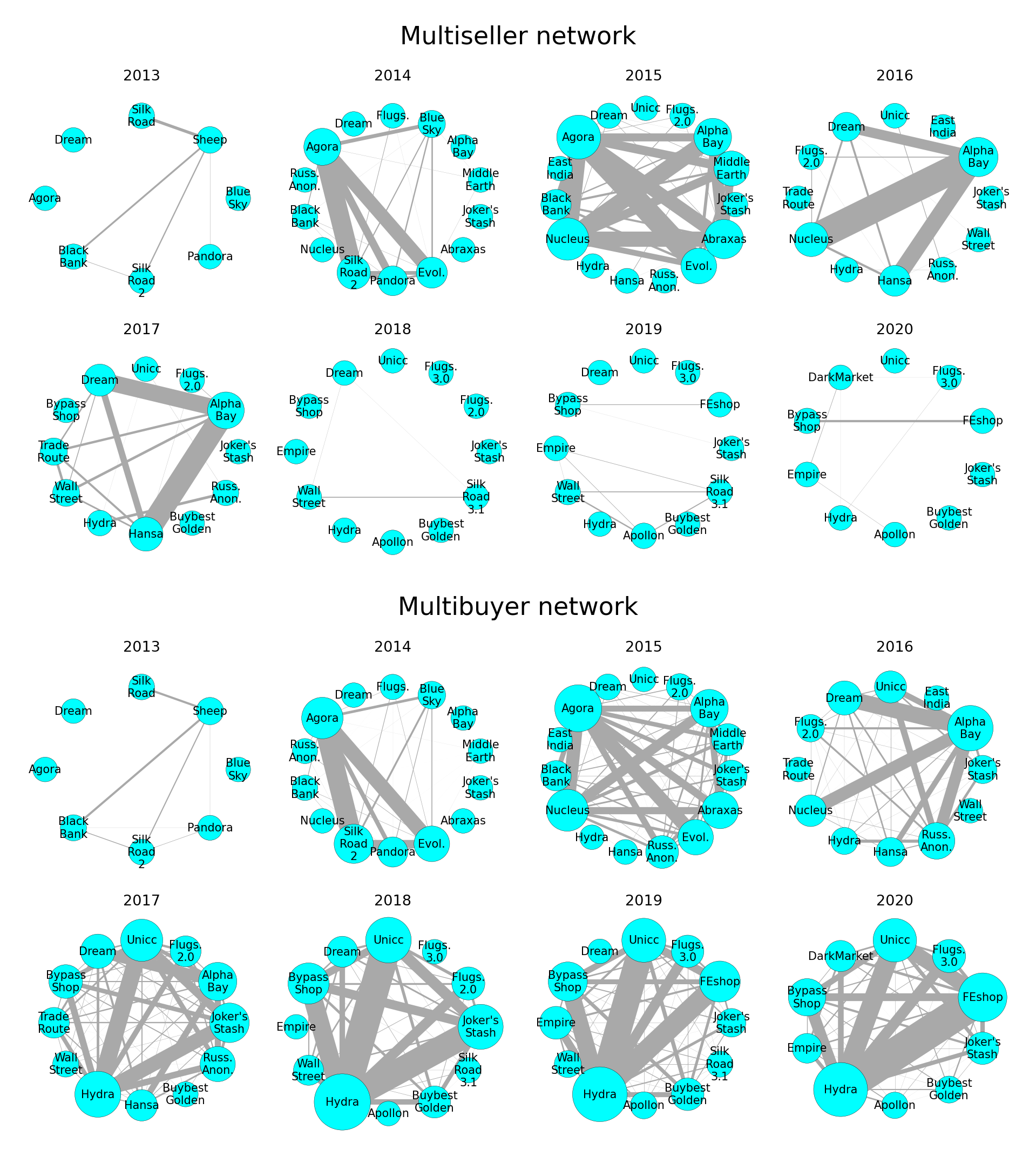}
	\caption{
		\textbf{The structural change in the multiseller network and the resilience of the multibuyer network.}
	    Temporal network of multisellers (top) and multibuyers (bottom) between markets for each year.
		Nodes are markets that are active during the year.
        Edges are multihomers, i.e., traders that are simultaneously active in both markets (sellers in multiseller network, and buyers in the multibuyer network).
		The width of the edges is proportional to the number of multihomers acting between the markets.
		}
	\label{fig:multiseller_multibuyer_network}
\end{figure*}

To investigate the role of sellers on the broader ecosystem of DMWs sellers, we consider a temporal network where nodes are the active markets, and an edge between two nodes exists when a multiseller is present between them, what we henceforth call the \textit{multiseller network}. 
The evolution of the multiseller network is shown in Figs.~\ref{fig:multiseller_multibuyer_network}.
Until 2012, there is only one active market, namely Silk Road market, and hence no multihomer activity.
From 2013 until 2015, the multiseller network grows in terms of connectivity, showing an increasing number of edges spread across different markets.
During 2016 and 2017, the edges are polarized by AlphaBay, the dominant market.
Then, between 2017 to 2018, there is a drastic structural change in the multiseller network structure due to operation Bayonet, after which the connections almost vanished.
This change persists until the end of the observed period of the data set.

Although the number of multisellers suffers a severe drop [Fig.~\ref{fig:categories}(a)] and the multiseller network drastically shrinks (Fig.~\ref{fig:multiseller_multibuyer_network}), the net income of multisellers remains persistently the largest among sellers throughout the whole period of observation, as shown in Fig.~\ref{fig:categories}.
This suggests that the multiseller activity is sensitive to external shocks but also that it yields higher profits.

\subsection*{Multibuyer network} \label{sec:multibuyer_network}

Buyers simultaneously active on multiple markets also play the role of connectors in the ecosystem.
Therefore, we analyse the temporal network where nodes are the active markets and an edge between two nodes exists when a multibuyer is present between them, what we henceforth call the \textit{multibuyer network}.
The structural change seen in the multiseller network is not observed in the multibuyer network, as show in Fig.~\ref{fig:multiseller_multibuyer_network}.
The evolution of the multibuyer network follows a similar pattern to the multiseller network until 2015, despite a stronger polarization around Hydra instead of AlphaBay during 2017.
However, after the operation Bayonet, although the network shows a decrease in connectivity, it still remains highly connected and with a large number of active multibuyers.
Moreover, the network is already fully recovered in 2019 showing a strong resilience against external shocks.

\subsection*{Seller-to-seller network} \label{sec:s2s_network}

In order to investigate the role of direct transactions between market participants, we now analyse the evolution of the S2S network, i.e., the network of the U2U transactions involving only sellers.
The nodes of the S2S network are active sellers and two sellers are connected by an edge if at least one transaction was made between them during the considered snapshot period.
Although the S2S network is composed only of U2U transactions, all categories of sellers (i.e, market-only, U2U-only, and market-U2U) are present in the S2S network.
For instance, market-only sellers are entities classified as sellers only in markets, but that may promote U2U transactions with other sellers, therefore being part of the S2S network.
Therefore, the S2S network can be seen as a proxy for a distribution network of illegal products.

To reduce the presence of noise in the S2S network, we consider only stable U2U pairs, i.e., pairs that have at least three transactions throughout the whole period of observation \cite{Nadini2022SciRep}.
The traded volume generated by stable pairs is more than five times larger than that of non-stable pairs \cite{Nadini2022SciRep}.
The S2S network is mostly populated by U2U-only vendors, followed by market-only, and market-U2U (see Supplementary Information).

In Fig.~\ref{fig:v2v_network_bayonet}, we show the giant component of the S2S network one year before the operation Bayonet and one year after.
The network shows a notable structural change, significantly shrinking.
However, the evolution of the S2S network shows a different pattern than that observed in both the multiseller and the multibuyer networks.

\begin{figure*}[ht]
	\centering
	\includegraphics[scale=1]{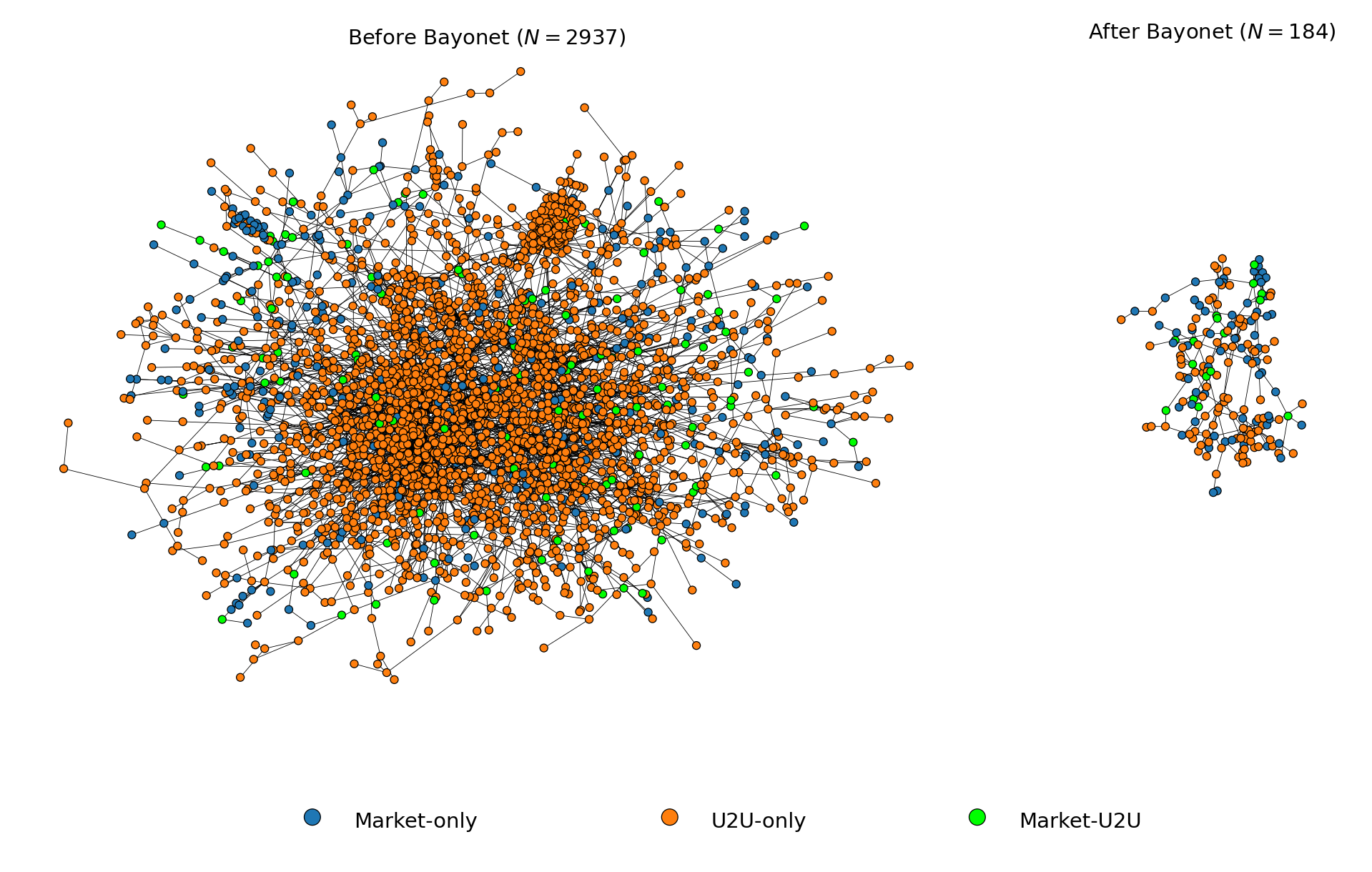}
	\caption{
        \textbf{The impact of the operation Bayonet on the S2S network.}
		The giant component of the  S2S network one year before and one year after the operation Bayonet.
		Nodes are sellers that are active within the time period, and an edge is placed between two sellers if at least one transaction occurs between them during the period.
		The S2S network is mostly populated by U2U-only sellers.
		}
	\label{fig:v2v_network_bayonet}
\end{figure*}

\begin{figure*}[ht!]
	\centering
	\includegraphics[scale=1]{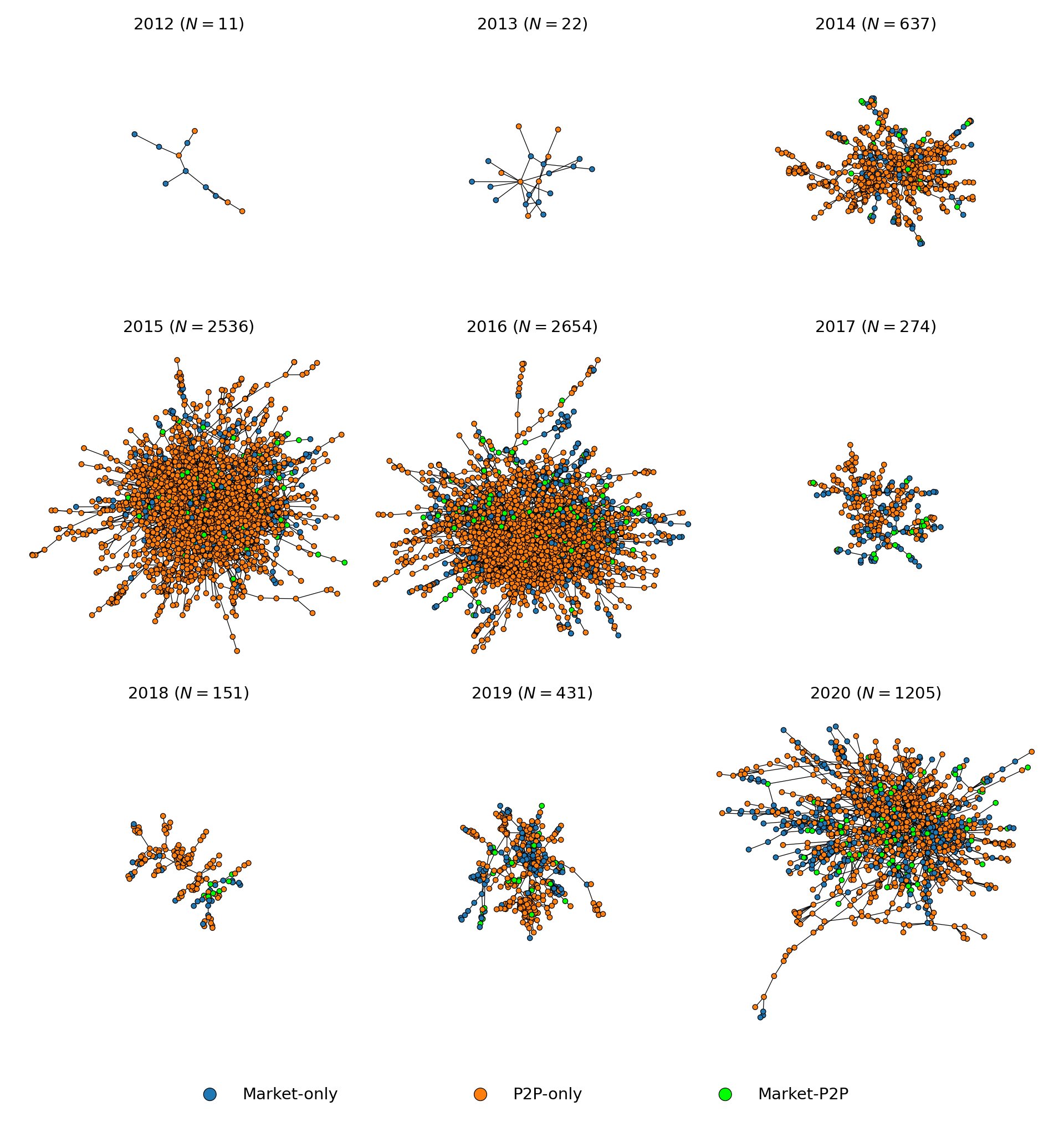}
	\caption{
        \textbf{The intermediate resilience regime of the S2S network.}
		The giant component of the S2S network of U2U transactions between sellers for each year with the respective number of nodes ($N$).
		The nodes are sellers that are active in that year, and an edge is placed between two sellers if at least one transaction occurs between them during that year.
		The network is mostly populated by U2U-only sellers, followed by market-only sellers.
		After a major external shock in 2017, the S2S network shrinks but, unlike the multiseller network, recovers, growing again, although not as fast as the multibuyer network.
		}
	\label{fig:s2s_network}
\end{figure*}

From 2012 to 2016, the giant component of S2S network continuously grows in number of nodes and connections, as shown in Figs.~\ref{fig:s2s_network}.
Then, during 2017 and 2018, it shows the structural change due to operation Bayonet, when it shrinks.
However, unlike the multiseller network, the S2S network recovers during 2019 and 2020, but slower than the multibuyer network recovery.
Therefore, the S2S network appears to be more resilient than the multiseller network but less than the multibuyer network.
The same pattern is observed in the whole S2S network (see Supplementary Information).

\section*{Methods} \label{sec:methods}

\subsection*{Data preprocessing} \label{sec:data_preprocessing}

Although in essence similar to usual online marketplaces, DWMs are illegal unregulated commercial websites.
In fact, despite technical differences, human behaviour in DWMs closely resembles that observed in usual regulated platforms \cite{Bracci2022PNASNexus}.
Nevertheless, due to their unregulated nature, DWMs exhibit behaviours not observed in regulated marketplaces.
They offer anonymity to their users by using and developing specialized tools.
DWMs are accessed through darknet browsers supporting the onion routing protocol (e.g., Tor), which provides anonymous communication connections  \cite{Reed1998Onion}.
Additionally, transactions are made with cryptocurrencies, mostly Bitcoin, which also provide anonymity to the transaction parties \cite{Nakamoto2008Bitcoin, Interpol2020CombCyberCrime}.
While the Bitcoin blockchain is publicly available on Bitcoin core \cite{Bitcore} or other third-party APIs such as \textit{Blockchain.com} \cite{Blockchain.com}, a market or a user can generate a new address for each transaction.
To track the transactions of markets and users as entities, the data need to be pre-processed in order to map group of addresses into entities.

We use data of DWM transactions on the Bitcoin blockchain pre-processed by Chainalysis Inc.
Although other coins are used, such as Monero recently, Bitcoin is still the mostly used in the ecosystem, being supported by more than 93\% of markets \cite{Europol2021IOCTA, Chainalysis2022CrimeReport}.
The pre-processing relies on established state-of-the-art heuristics to cluster addresses into entities, such as cospending, intelligence-base, and behavioral clustering \cite{Tasca2018evolution, Ron2013book, Meiklejohn2013InternetConf, Harrigan2016unreasonable}.
The resulting data set includes for each transaction the source and destination entities, the time, and the value of the transaction.

\begin{figure*} [ht!]
	\centering
	\includegraphics[scale=0.95]{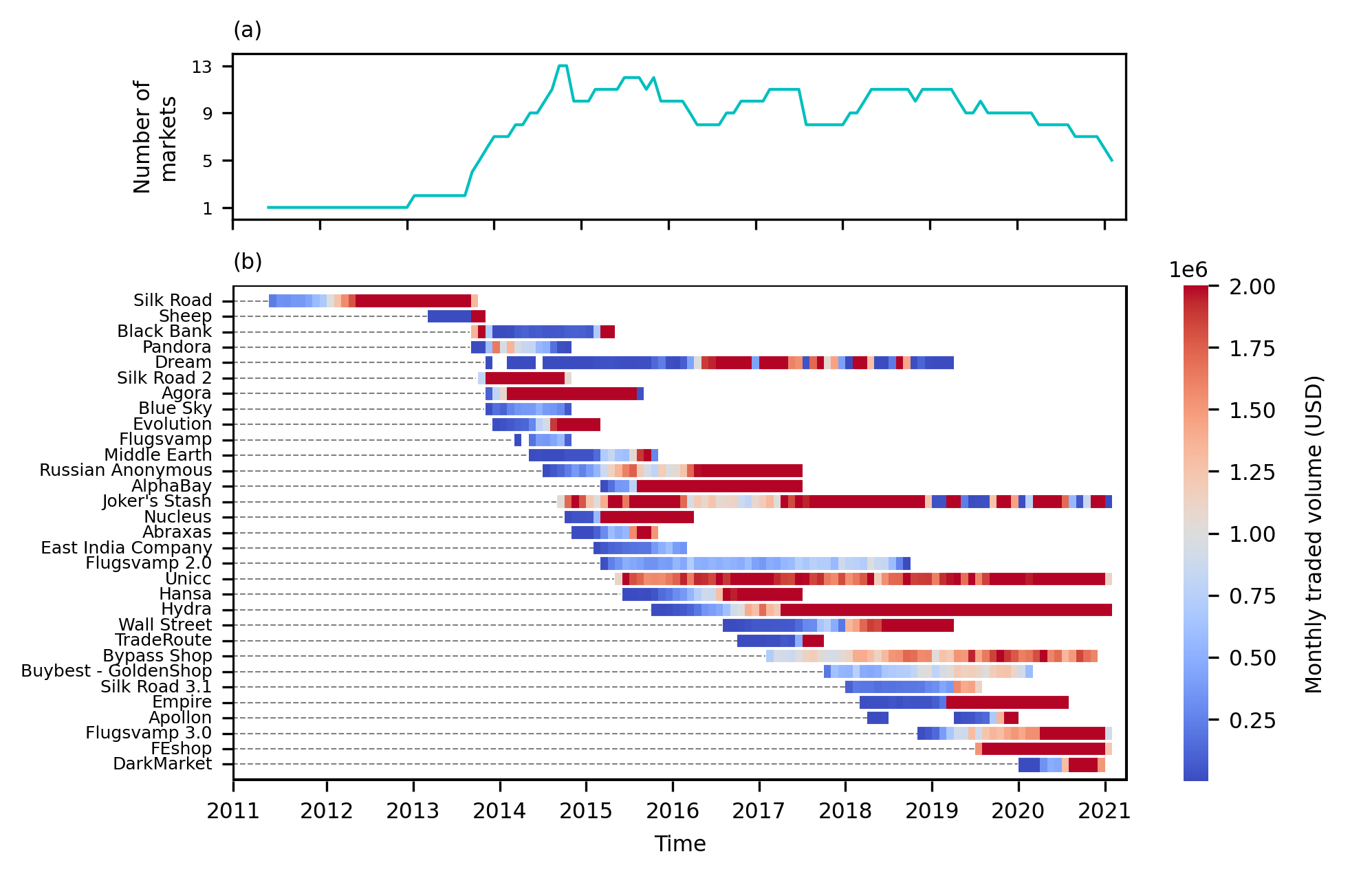}
	\caption{
		\textbf{The evolution of the ecosystem of DWMs.}
		(a) Total monthly number of active markets.
		(b) Lifetime of markets.
		The horizontal bars represent each market lifetime, i.e., the time when the market becomes active until its closure, and is colored according to the market's monthly traded volume in USD.
		In the vertical axis, markets are in the chronological order of their launch date, although for some markets the activity effectively starts after the launch date (e.g., AlphaBay).
		}
	\label{fig:volume_lifetime_markets}
\end{figure*}

We exclude transactions with a value larger than $5 \times 10^5$ USD and smaller than $0.01$ USD
To include the major marketplaces and obtain statistically relevant measures, we selected markets with an average daily traded volume larger than 15,000 USD and a lifetime larger than six months.
As a result, our data set consists of more than a decade of the entire transaction history of 31 DWMs between June 2011 and February 2021, as shown in Fig.~\ref{fig:volume_lifetime_markets}.

\subsection*{Network structure of transactions} \label{sec:network_structure}

We represent the network of transactions by temporal networks where nodes are entities (markets and users), and directed edges represent a transaction pointing from the source to the destination entity and endowed with the time and value of the transaction.
Each marketplace is a star-graph where the central node is the marketplace, and the leaf nodes, i.e., the first-neighbors, are the marketplace users.
Therefore, all transactions involving the market have the market either as a source or as a destination node.

Additionally, we analyse the U2U network of transactions, i.e., the transactions between pairs of market first-neighbors where the source and destination nodes are market users without the market as an intermediate.
In the U2U network, an edge connects nodes that are not necessarily users of the same market.
Therefore, the U2U network connects different market star-graphs.
Previous studies have shown that, although the number of users and transactions is larger in markets, the traded volume in the U2U network is larger than that of markets \cite{Nadini2022SciRep}.

\subsection*{Classification of sellers and buyers} \label{sec:classification}

We classify all entities either as sellers or buyers as a function of time.
The result is a time series of lists of sellers and buyers for each period and for each market and the U2U network.
The classification is performed in five steps (see Fig.~\ref{fig:classification_steps}), as detailed next.

\textbf{Step 1: Intersection of features.}
We use six features to classify sellers: (1) money received, (2) ratio between received and sent money, (3) number of received transactions, (4) ratio between number of received and sent transactions, (5) lifetime, and (6) mean interevent time of transactions, as detailed next.

The classification is a function of time.
For each entity, we keep track of the cumulative values of the six features for each transaction performed by the entity.
In other words, we consider each of the six features cumulatively as a function of time.
Then, we assign a threshold value for each feature.
To be classified as a seller, an entity must simultaneously satisfy the six feature threshold criteria, as follows.

First, we consider the total cumulative received money by each entity $i$ at time $t$, denoted by $m^{\rm rec}_i(t)$.
This feature is satisfied when $m^{\rm rec}_i(t) \geq M$, where $M$ is the fixed value of minimum total cumulative received money.

Second, we consider the ratio between the total cumulative received and sent money by each entity $i$ at time $t$, denoted by $R_i(t)$, i.e., $R_i(t) \equiv m^{\rm rec}_i(t) / m^{\rm sent}_i(t)$, where $m^{\rm sent}_i(t)$ is the total cumulative sent money by entity $i$ at time $t$.
This feature is satisfied when $R_i(t) \geq \alpha$, where $\alpha$ is a constant value.

Third, we consider the total cumulative number of received transactions by each entity $i$ at time $t$, denoted by $n_i^{\rm rec}(t)$.
This feature is satisfied when $n^{\rm rec}_i(t) \geq T$, where $T$ is the fixed value of minimum total cumulative number of received transactions.

Fourth, we consider the ratio between the total cumulative number of received and sent transactions by each entity $i$ at time $t$, denoted by $Q_i(t)$, i.e., $Q_i(t) \equiv n^{\rm rec}_i(t) / n^{\rm sent}_i(t)$, where $n^{\rm sent}_i(t)$ is the total cumulative number of transactions sent by entity $i$ at time $t$.
This feature is satisfied when $Q_i(t) \geq \beta$, where $\beta$ is a constant value.

Fifth, we consider the lifetime of each entity $i$ at time $t$, defined as the time interval between the first and the last transaction performed by the entity until time $t$, denoted by $\ell_i(t)$.
This feature is satisfied when $\ell_i(t) \geq L$, where $L$ is the fixed value of minimum lifetime.

Sixth, we consider the cumulative mean interevent time for each entity $i$ at time $t$, defined as the mean of the sequence of time interval between consecutive transactions of an entity until time $t$, which we denote by $\phi_i(t)$.
This feature is satisfied when $\phi_i(t) \leq \tau$, where $\tau$ is the maximum value of mean interevent time.

Finally, given a set of parameters $\{M, \alpha, T, \beta, L, \tau\}$, we classify an entity $i$ as a seller if 
$m^{\rm rec}_i(t) \geq M$, 
$R_i(t) \geq \alpha$, 
$n^{\rm rec}_i(t) \geq T$, 
$Q_i(t) \geq \beta$, 
$\ell_i(t) \geq L$, and 
$\phi_i(t) \leq \tau$ are simultaneously satisfied at time $t$.
Conversely, because the classification is a function of time, an entity that has been classified as a seller at time $t$, might lose the seller status if, at a future time, the features are not satisfied.
This step is performed separately for each market and the U2U network.
After classifying each entity according to its time series of transactions, we aggregate sellers daily, i.e., we obtain a daily time series of lists of sellers for each market and the U2U network.

\textbf{Step 2: Sliding time window.}
The method used in step 1 captures the activity of entities in a continuous-time framework, i.e., the features are computed for each transaction taken by each entity.
This results in a time series of sellers where sellers are irregularly classified because of oscillations on each entity specific activity, such as having a less frequent number of transactions during a period.
Therefore, we use a sliding window of $\Delta t$ days to classify sellers, i.e., every day that an entity is classified as a seller, it remains as a seller for $\Delta t$ days, including the first day.
After using the sliding time window for sellers, all entities that are not classified as sellers are classified as buyers for each day.
At the end of step 2, we generate a daily time series of sellers and buyers for each market and the U2U network.

\textbf{Step 3: Multihomers.}
The daily time series of sellers obtained in step 2 is a list of sellers for each day for each market.
To identify multisellers, we first compute, for each pair of simultaneously active markets, the intersections of the daily lists of sellers obtained from step 2.
Then, to obtain the daily time series of multisellers, we compute the union of the daily intersections of sellers between pairs of markets.
We perform the same procedure to compute the daily time series of multibuyers but using the daily time series of buyers obtained from step 2.

\textbf{Step 4: Aggregation by period.}
To observe the behavior of the ecosystem on specific calendar periods, such as weekly or quarterly, we select a time period and aggregate the daily time series through step 3 accordingly.
For example, to obtain a monthly time series of sellers, we compute the union of the lists of sellers for each month.
This step is independent of the sliding time window in step 2.
For instance, if an entity is classified as a seller for 20 consecutive days in a month in Step 1 and $\Delta t = 30$ days, that entity will remain as a seller for 30 days from the last day of the 20 days, hence still being a seller in the next month.
Therefore, at the end of step 4, we obtain a time series of buyers and sellers for each market and the U2U network according to the selected time period.

\textbf{Step 5: Categories.}
For each period of time obtained in step 4, some sellers are active only in markets, others in the U2U network, or in both.
Therefore, for each time period, we divide the sellers into three mutually exclusive categories: (1) market-only sellers, which are the union of sellers that are active in one or more markets but only markets and not in the U2U network; (2) U2U-only sellers, which are the union of sellers that are active only in the U2U network; and (3) market-U2U sellers, which are the union of sellers who are active in one or more markets and also active in the U2U network.
For instance, multisellers belong to set of market-only or market-U2U sellers, but not to the set of U2U-only sellers by definition.
Analogously, we divide buyers for each time period into three mutually exclusive categories: market-only buyers, U2U-only buyers, and market-U2U buyers.
Specifically for buyers, when we compute the union or intersection of sellers across markets and the U2U network, we remove entities that are sellers in any market or the U2U network in that time period.

\section*{Discussion} \label{sec:discussion}

In this paper, we proposed a method for classifying users as sellers or buyers in the ecosystem of DWMs.
We then identified three key categories of buyers and sellers that play different roles in the ecosystem: Market-only traders that are active only in markets and not in the U2U network; U2U-only traders that are active only in the U2U network; and market-U2U traders that are active on both markets and the U2U network simultaneously.
Additionally, we singled out the multihomers, i.e., users that are simultaneously active in multiple markets, acting either as sellers (the multisellers), or as buyers (the multibuyers). 

We showed that a small fraction of traders concentrate a large fraction of the traded volume, and by analysing the networks of buyers and sellers, we found different resilience regimes.
Shocks tend to induce serious structural changes in the multiseller network, but impact much less severely the multibuyer network.
Interestingly, the S2S network shows an intermediate resilience, which can be seen as a hint that the S2S network might play the role of a supply chain network of sellers.
Furthermore, after a shock, the activity of buyers is resumed almost immediately, while sellers recover more slowly.
These different regimes suggest that the ecosystem's resilience is mainly supported by the high demand of buyers that the reactivity of sellers.

Despite consistent results, this study has limitations that may be addressed in future work.
First, while the dataset is preprocessed with  state-of-the-art methods, there is no ground truth for validation, and this uncertainty propagates to our findings.
For instance, we cannot verify if an entity classified as seller is in fact a seller.
Similarly, there is no unique choice for the classification parameters or ground truth for fitting them.
Nevertheless, it is important to stress that the results are robust under considerable variation of the parameters, indicating that the coherent picture emerging from our analysis does not depend on the details of the method. Future work may further extend the approach presented here, for example using machine learning methods to capture further behavioral regularities.
Second, our approach does not explicitly classify buyers, which are entities that were not classified as sellers. 
Despite coherent results, this clearly leaves space for refinements.
Third, at any given moment we classify entities as either buyers or sellers.
Yet it is possible that multiple roles are played at once. 
For example, in some cases, a seller in a given market may behave as a buyer in second market or in the U2U network. 
This multi-role classification, to be implemented in future work, can help gain a more nuanced understanding of the ecosystem.

Overall, our study provides a first step towards a better microscopic characterisation of the DWM ecosystem, indicating a direction of investigation that may be of interest to both researchers and law enforcement agencies. 
The results further support the recent efforts of law enforcement agencies to focus on individual sellers \cite{Horton2021GlobalDrugPolObserv, DOJ2018Operation, Europol2020DisrupTor}, as well as, more recently, also buyers \cite{Europol2021DarkHunTor, Europol2023SpecTor}.
The finding that multihomers and, in specific cases, multibuyers play a central role in connecting the ecosystem, thus contributing to its resilience, may indicate on how to better target future law enforcement operations. 
In general, by understanding the operation of key players within the DWM ecosystem, our work highlights how appropriate strategies can be designed to counteract the online trade of illicit goods more effectively.

\section*{Acknowledgements}
The research was partially supported by the Alan Turing Institute, also thanks to a generous gift from PayPal.

\section*{Data availability}
All data needed to evaluate the conclusions in the paper are present in the paper. Additional data related to this paper may be requested from the authors.

\bibliographystyle{unsrtnat} 
\bibliography{ref_dwm_multihomers}

\newpage
\onecolumn

\vspace*{1cm}
\begin{center}
    \textbf{\LARGE Supplementary Information}
\end{center}
\vspace{0.5cm}

\section*{Robustness of classification} \label{sec:classification_robustness}

Our classification method has seven parameters, namely the set of six features for classifying sellers, $\{M, \alpha, T, \beta, L, \tau\}$, and the size of the sliding time window $\Delta t$.
To test the robustness of the classification with respect to the parameters, we run stress-tests where we vary one of the  parameters independently and keep the other six fixed.
In the following, when varying one of the parameters, we set the other parameters with the same values used in the main text, i.e., $M = 100$ USD, $\alpha = 3$, $T = 10$, $\beta = 3$, $L = 10$ days, $\tau = 10$ days, and $\Delta t = 30$.
The classification under these values generates basal sets of buyers and sellers that are already conservative.
Therefore, by varying the parameter values under these strict conditions, we induce extreme cases for the robustness stress-tests.

The set of buyers is effectively unaffected by the change of parameters, unless extreme values are used.
For instance, if we use negative values in the first step of the classification.
Therefore, we focused on variations caused in the set of sellers.

We performed robustness tests on markets and the U2U network separately because they have different aspects.
For instance, a seller in the U2U network is a single vendor shop, while a seller in a DWM is part of a marketplace with many sellers.
Most importantly, the time frame of the U2U network is the whole observation period, while each market has its own time frame which is the market lifetime hence much shorter than the U2U network time frame.

\begin{figure} [ht!]
	\centering
	\includegraphics[scale=1]{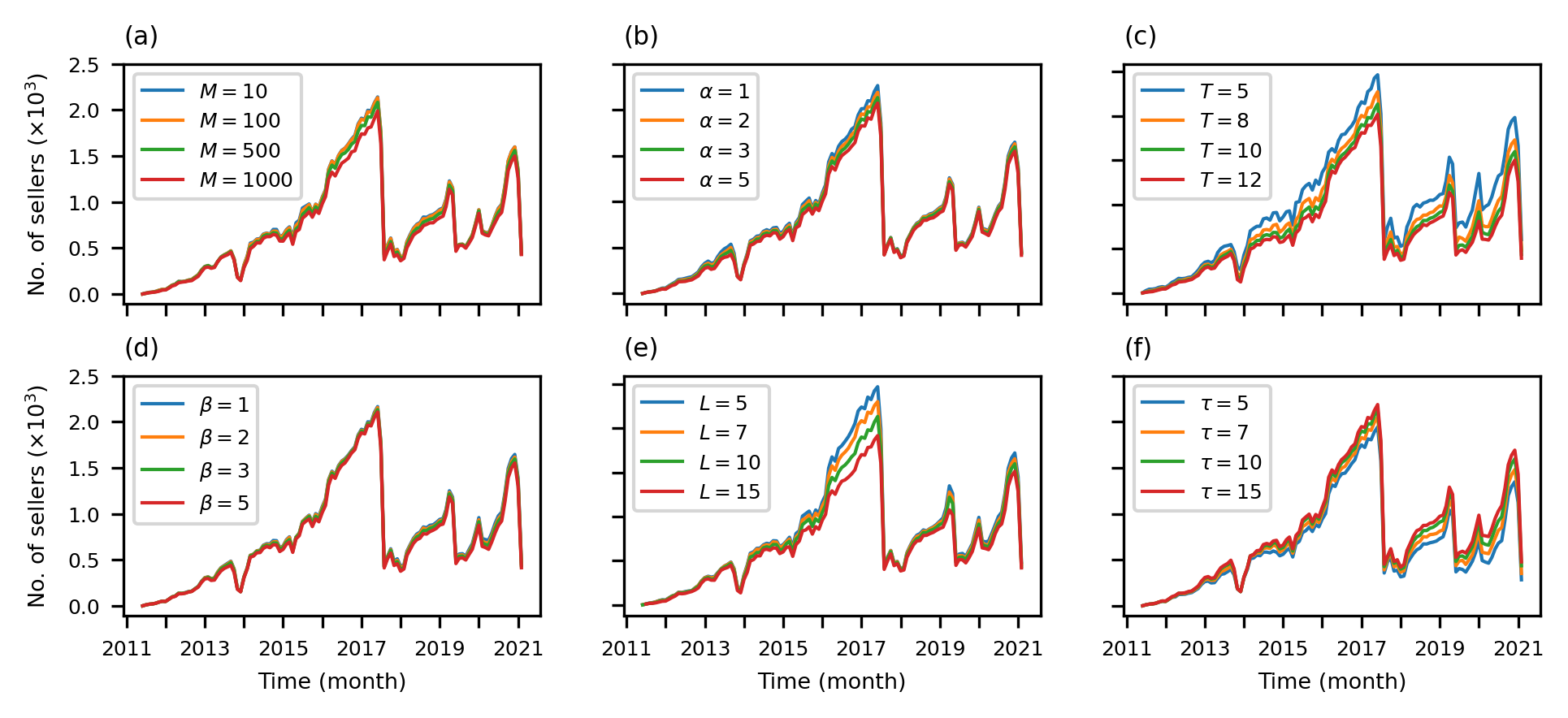}
	\caption{
		\textbf{Robustness of the classification with respect to the six features for markets}.
		We vary each feature independently (i.e., we vary one feature and keep the others fixed) and run the whole classification for all markets.
		(a) Obtained number of sellers when $M = $ 10, 100, 500, and 1000 USD.
		(b) Obtained number of sellers when $\alpha = $ 1, 2, 3, 4, and 5.
		(c) Obtained number of sellers when $T = $ 5, 8, 10, and 12 transactions.
		(d) Obtained number of sellers when $\beta = $ 1, 2, 3, and 5.
		(e) Obtained number of sellers when $L = $ 5, 7, 10, and 15 days.
		(f) Obtained number of sellers when $\tau = $ 5, 7, 10, and 15 days.
		}
	\label{fig:SI_robustness_market_features}
\end{figure}

First, we analyse each of the six features used for classifying sellers in step 1 of the classification (see Methods in the main text).
In Figure \ref{fig:SI_robustness_market_features}, we show the obtained time series of sellers for all markets when varying each of the six features.
Overall, the results show that the classification is robust with respect to the features for markets, yielding the same qualitative results with a relative small variation.
Specifically for the minimum number of transactions (Fig.~\ref{fig:SI_robustness_market_features}(c)), the classification shows a larger variation compared to the other features because it selects more sellers for a small number of $T$.
We also observe a larger variation during 2016 and 2017 when varying the minimum lifetime (Fig.~\ref{fig:SI_robustness_market_features}(e)).
During this period, AlphaBay is the largest market which may attract several market sellers but that do not stay active for a long period.

\begin{figure} [ht!]
	\centering
	\includegraphics[scale=1]{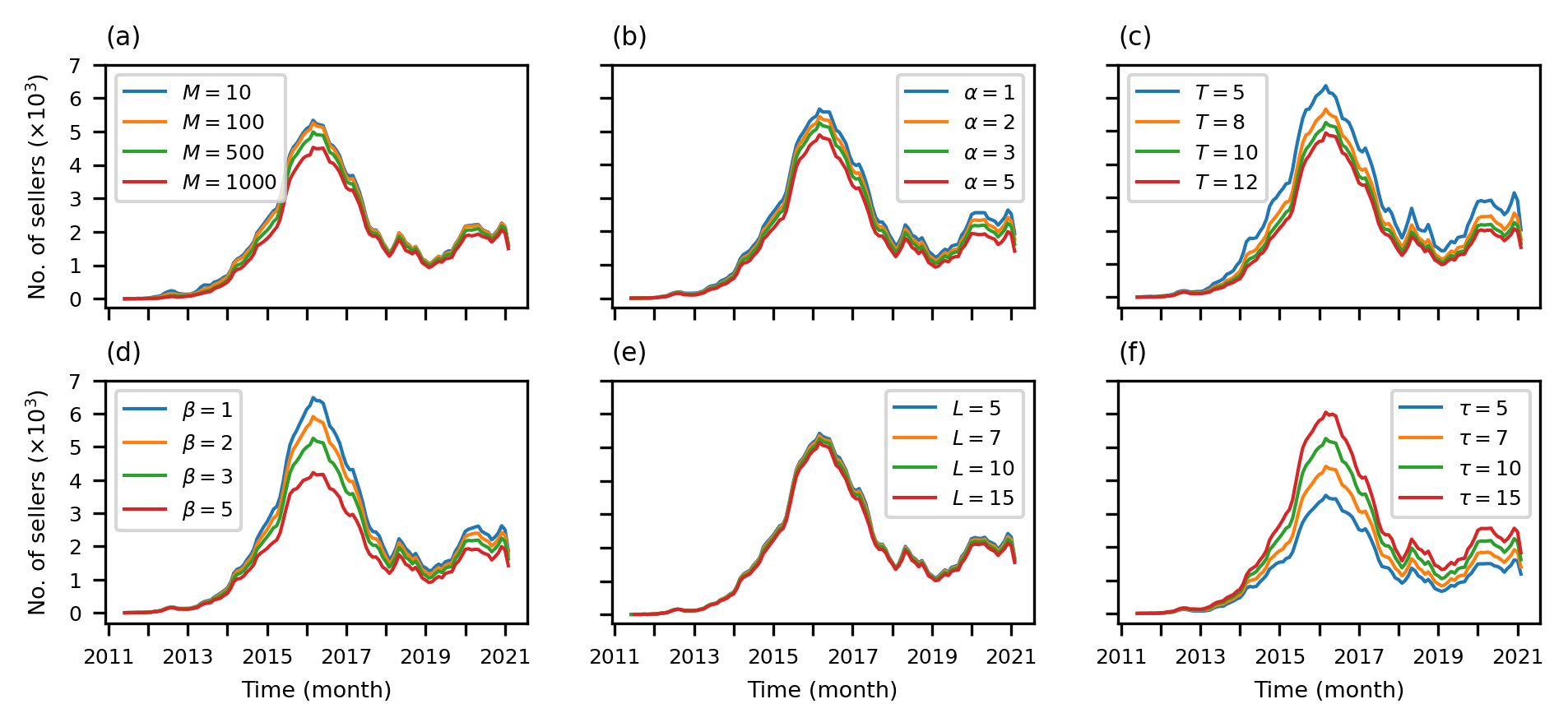}
	\caption{
		\textbf{Robustness of the classification with respect to the six features for the U2U network}.
		We vary each feature independently (i.e., we vary one feature and keep the others fixed) and run the whole classification for the U2U network.
		(a) Obtained number of sellers when $M = $ 10, 100, 500, and 1000 USD.
		(b) Obtained number of sellers when $\alpha = $ 1, 2, 3, 4, and 5.
		(c) Obtained number of sellers when $T = $ 5, 8, 10, and 12 transactions.
		(d) Obtained number of sellers when $\beta = $ 1, 2, 3, and 5.
		(e) Obtained number of sellers when $L = $ 5, 7, 10, and 15 days.
		(f) Obtained number of sellers when $\tau = $ 5, 7, 10, and 15 days.
		}
	\label{fig:SI_robustness_p2p_features}
\end{figure}

The classification is also qualitatively robust in the U2U network, as shown in Fig.~\ref{fig:SI_robustness_p2p_features}.
However, the variation is larger in the U2U than markets for $\beta$ and $\tau$.
Because the time frame of the U2U is the whole period of observation, it is more unlikely that users keep a small mean interevent time in the long term.
The large variation on $\beta$ may be associated to the fact that, compared to markets, there are less transactions in the U2U network, and U2U users may have a larger number of outgoing number transactions.
Therefore, one may find reasonable to set different values to the six features used for the U2U network than those used for markets.
However, we kept the same values across markets and the U2U network to consistently classify sellers, grasping the same behaviour in both situations.

\begin{figure} [ht!]
	\centering
	\includegraphics[scale=1]{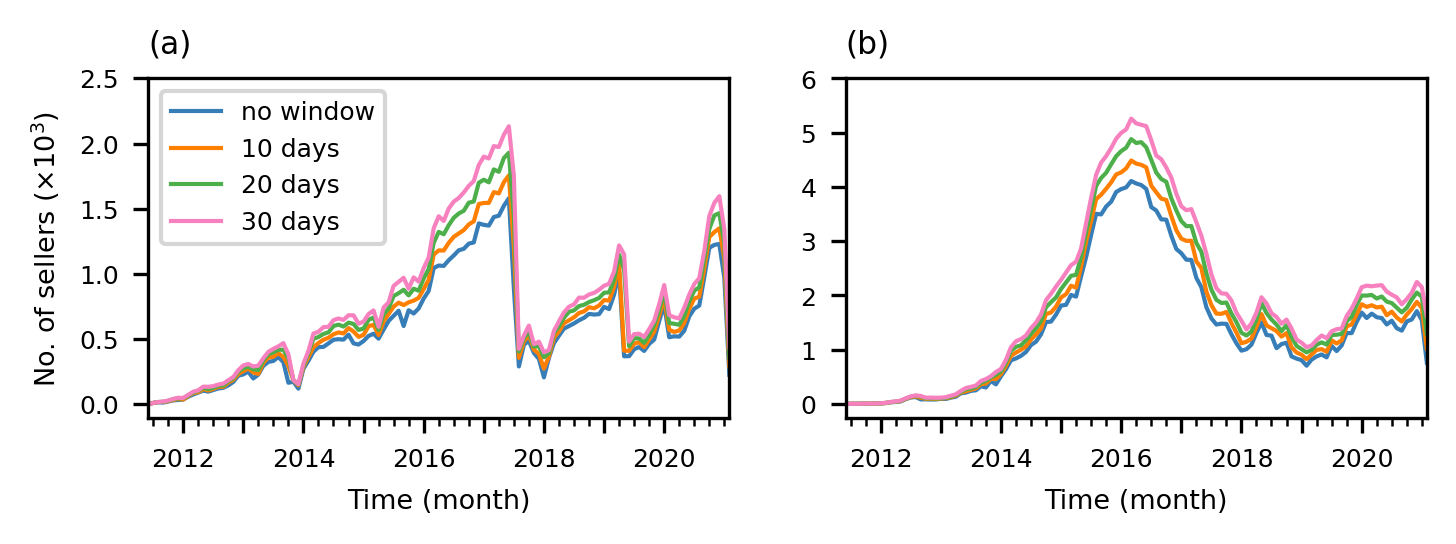}
	\caption{
		\textbf{Robustness of the classification of sellers with respect to the sliding time window, $\Delta t$}.
		The number of sellers when $\Delta t = $ 10, 20, and 30 days, and no sliding time window (no window) is used.
		(a) Sellers of all markets.
		(b) Sellers in the U2U network.
		}
	\label{fig:SI_robustness_sliding_window}
\end{figure}

Next, we analyse the size of the sliding time window, $\Delta t$, used in step 2 of the classification (see Methods in the main text).
We compare the results between no sliding time window and with $\Delta t = $ 10, 20, and 30 days, as shown in Fig.~\ref{fig:SI_robustness_sliding_window}.
The number of sellers increases as $\Delta t$ increases, as expected, but not significantly.
The results for different values of $\Delta t$ are qualitatively equivalent, showing that the classification is robust against the choice of $\Delta t$.

\begin{figure} [ht!]
	\centering
	\includegraphics[scale=1]{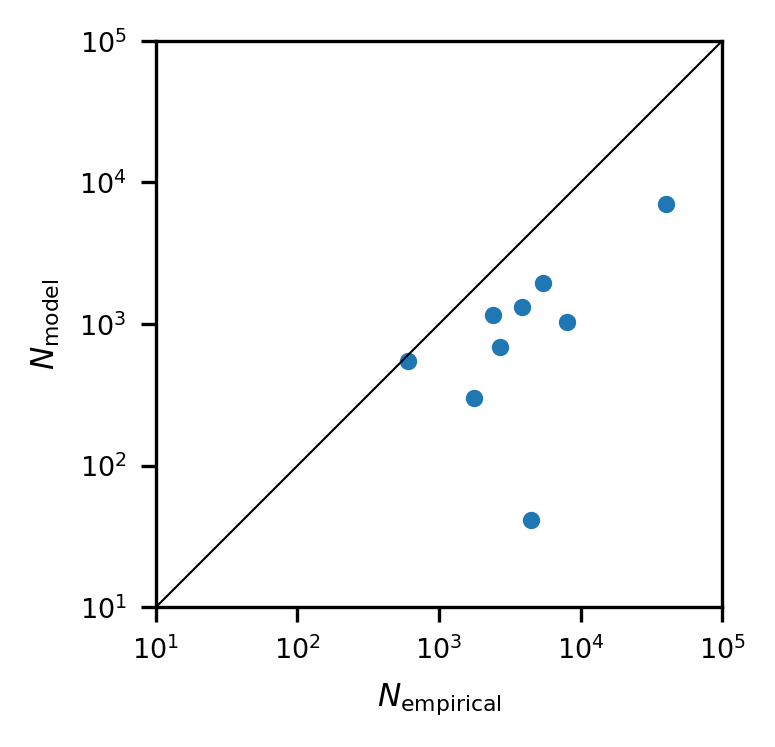}
	\caption{
		\textbf{Conservative classification parameters.}
        Comparison between the number of sellers of a market obtained from official sources ($N_{empirical}$) and obtained by our model ($N_{model}$).
		}
	\label{fig:validation}
\end{figure}

Finally, to validate our classification model against an empirical measure, we compare the total number of sellers obtained by our model for a market with the real number of sellers of that market.
We collect from different sources the number of sellers of nine markets as shown in table \ref{tab:validation}.
We conservatively chose the classification parameter values, such that the numbers of sellers obtained by our model underestimate the empirical numbers, as shown in Fig.~\ref{fig:validation}.
Moreover, although the robustness tests show that the number of sellers change according to the choice of parameters, our results are qualitatively unchanged.

\begin{table}[ht]
   \caption{\label{tab:validation}
   The number of sellers of each market and the source.
	}
   \begin{tabular}{ccc}
     \textbf{Market}     & \textbf{Number of sellers}  &  \textbf{Source}  \\ 
     \hline 
    \cmidrule{1-3}
          Silk Road      &            3877             &    \cite{SilkRoadTrial} \\
    \cmidrule{1-3}
          Evolution      &            2702             &      \cite{Rhumorbarbe2016ForensicSciInt} \\
    \cmidrule{1-3}
          AlphaBay       &            40000            &    \cite{DOJ2017AlphaBay}  \\
    \cmidrule{1-3}
          Flugsvamp 2.0  &            600              &     \cite{Presseportal2019Flugsvamp2} \\
    \cmidrule{1-3}
          Hansa          &            8000             &     \cite{NetherlandsPolice2017Hansa} \\
    \cmidrule{1-3}
          Wall Streeet   &            5400             &     \cite{Europol2019WallStreet} \\
    \cmidrule{1-3}
          Empire         &            4500             &     \cite{Webz.io2020Empire} \\
    \cmidrule{1-3}
          Apollon        &            1761             &     \cite{DarkOwl2020Apollon} \\
    \cmidrule{1-3}
          DarkMarket     &            2400             &     \cite{Europol2021DarkMarket} \\
   \end{tabular}
\end{table}

\section*{Composition of multisellers} \label{sec:multiseller_composition}

\begin{figure} [ht!]
	\centering
	\includegraphics[scale=1]{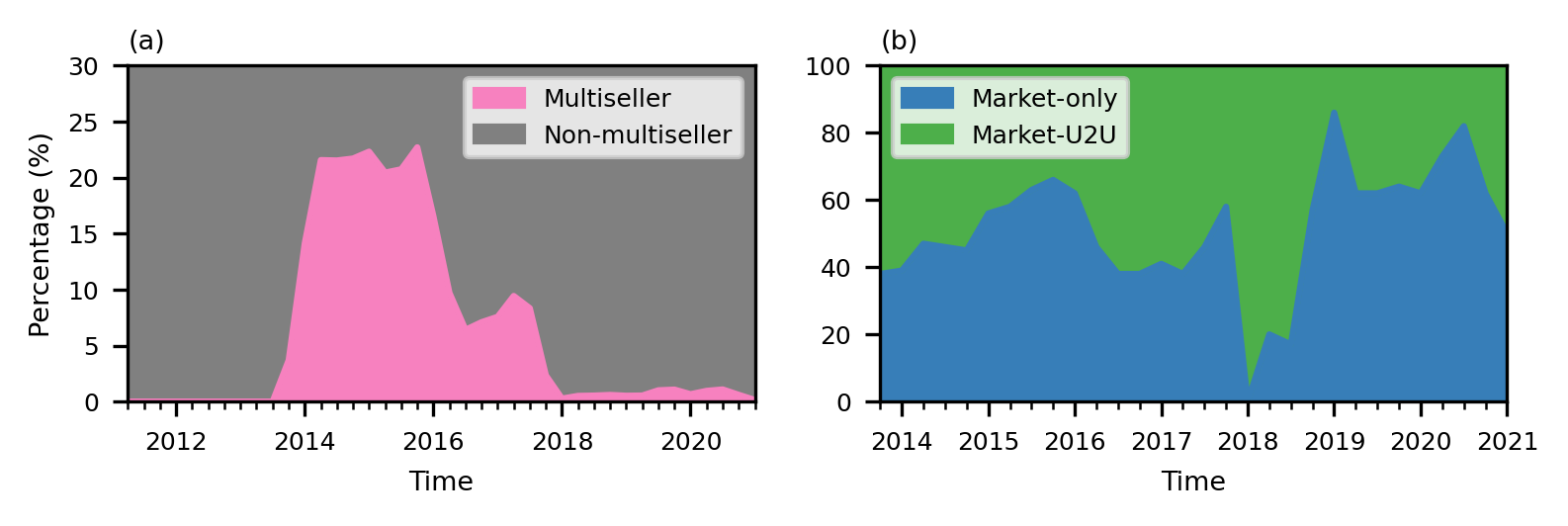}
	\caption{
		\textbf{Evolution of the composition of multisellers.}
        (a) Quarterly percentage of multisellers and non-multisellers among all sellers. 
        (b) Quarterly percentage of multisellers that are market-only sellers and market-U2U sellers. 
		}
	\label{fig:multiseller_composition}
\end{figure}

Multisellers start to appear by the end of 2013.
Their number increases fast, representing more than 20\% of all sellers until the first quarter of 2016, as shown in Fig.~\ref{fig:multiseller_composition}(a).
Then, until the end of 2017, their percentage decreases to about 10\% of all sellers.
After the operation Bayonet, multisellers represent less 3\% of sellers. 

By definition, multisellers are either market-only sellers or market-U2U sellers.
In Fig.~\ref{fig:multiseller_composition}(b), we show the composition of multisellers as percentages of market-only and market-U2U sellers.
Before 2018, the composition oscillates and multisellers are roughly composed by equal parts of the two categories of sellers.
During the first three quarter of 2018, more than 80\% of multisellers are market-U2U.
Then, from the last quarter of 2018 until the end of the observation period, multisellers are composed mostly by market-only sellers.

\section*{Composition and evolution of the S2S network} \label{sec:s2s_network_composition}

\begin{figure} [ht!]
	\centering
	\includegraphics[scale=1]{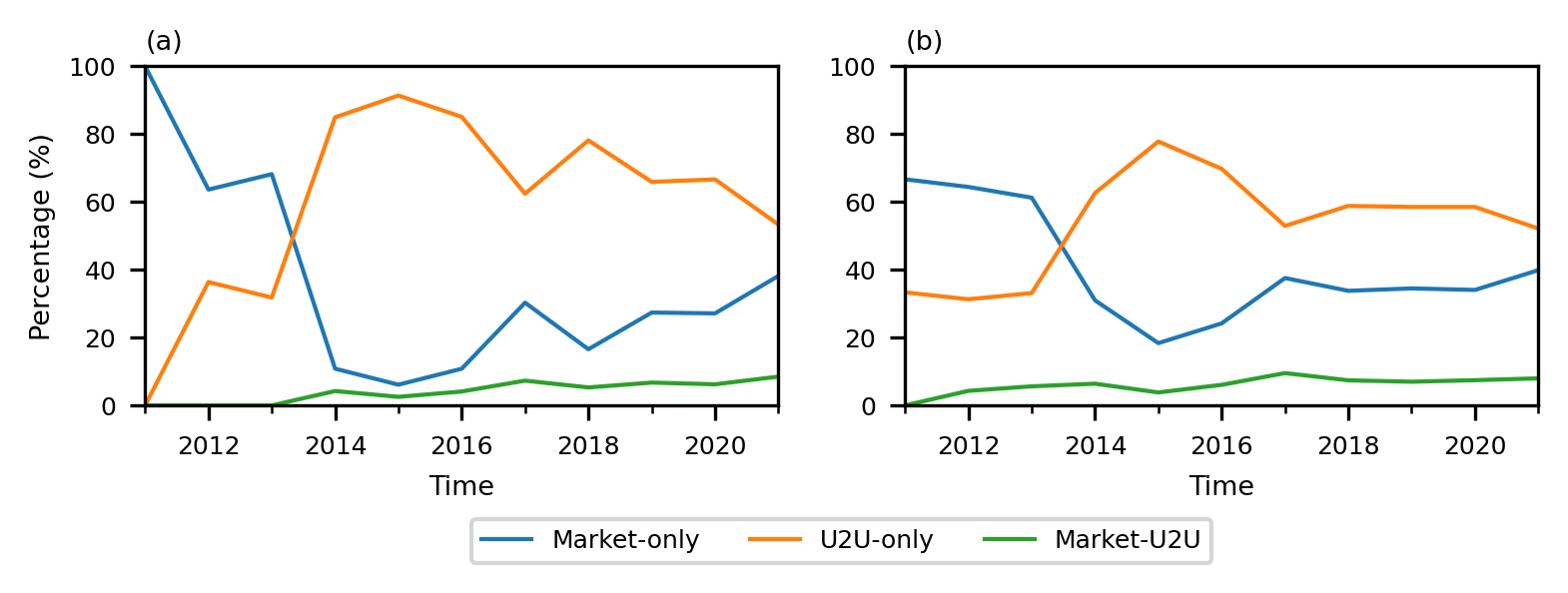}
	\caption{
		\textbf{Evolution of the S2S network composition.}
        Percentage of each category of seller in (a) the giant component of the S2S network and (b) the whole S2S network for each year.
		}
	\label{fig:s2s_composition_s30}
\end{figure}

\begin{figure*}[ht!]
	\centering
	\includegraphics[scale=1]{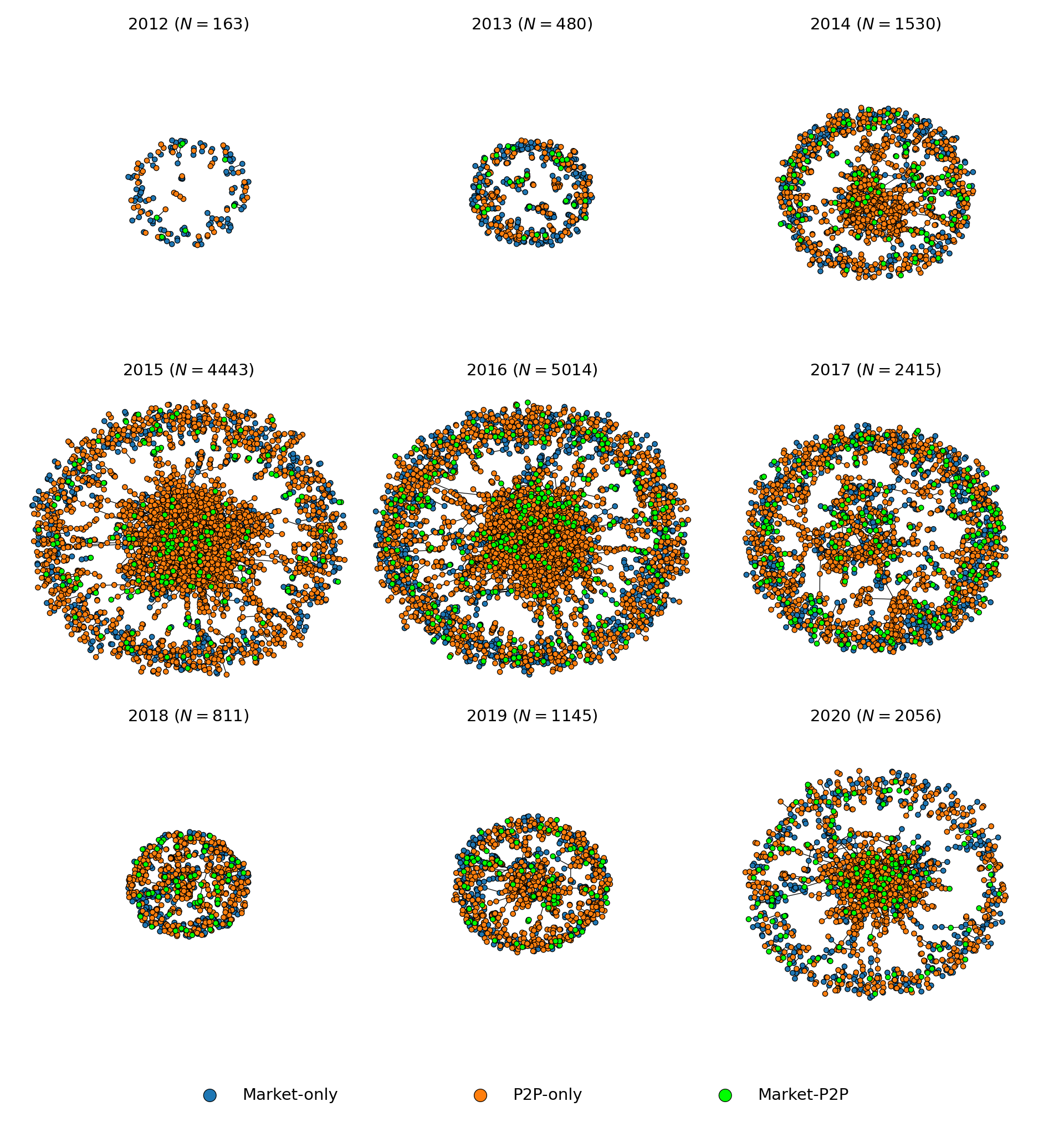}
	\caption{
        \textbf{Evolution of the S2S network.}
        The S2S network of U2U transactions between sellers for each year with the respective number of nodes ($N$).
		The nodes are sellers that are active in that year, and an edge is placed between two sellers if at least one transaction occurs between them during that year.
        The whole S2S network shows the same evolution pattern observed in the giant component (Fig.~\ref{fig:s2s_network}).
		}
	\label{fig:s2s_network_s30_stableP2P_whole1}
\end{figure*}

Until the beginning of 2013, when Silk Road is the dominant market, the giant component of the S2S network is composed mostly by market-only sellers, followed by U2U-only, and market-U2U sellers, as shown in Fig.~\ref{fig:s2s_composition_s30}(a).
Then, after that period until the end of our observation time, U2U-only is the dominant category in the giant component of the S2S network, followed by market-only, and market-U2U sellers.
The same pattern is observed in the whole S2S network, as shown in Fig.~\ref{fig:s2s_composition_s30}(b).
Moreover, the same evolution pattern observed in the S2S network giant component (Fig.~\ref{fig:s2s_network}) is also observed in the whole S2S network, as shown in Fig.~\ref{fig:s2s_network_s30_stableP2P_whole1}.

\end{document}